\documentclass[12pt,aps]{revtex4}

\usepackage{epsfig}
\usepackage{bbm}
\usepackage{amssymb}
\newcommand{\nc}{\newcommand}
\nc{\be}{\begin{equation}}
\nc{\ee}{\end{equation}}
\nc{\bea}{\begin{eqnarray}}
\nc{\eea}{\end{eqnarray}}
\nc{\nn}{\nonumber}

\nc{\acom}[2]{ \left\{ #1,#2 \right\} }
\nc{\com}[2]{ \left[ #1,#2 \right] }

\nc{\lp}{\left(}
\nc{\rp}{\right)}

\nc{\eq}{{Eq.}}
\nc{\eqs}{{Eqs.}}

\begin{document}
\rightline{CERN-PH-TH/2006-041}

%\vskip 0.2in

\title{Numerical Approach to Multi Dimensional Phase Transitions}

%\author{Thomas Konstandin}
%\address{Department of Theoretical Physics, Royal Institute of Technology (KTH), \\
%	AlbaNova University Center, Roslagstullsbacken 11, 106 91
%	Stockholm, Sweden}
%\ead{Konstand@kth.se}
%
%\author{Stephan J. Huber}
%\address{Theory Division, CERN,
%	CH-1211, Geneva 23, Switzerland}
%\ead{Stephan.Huber@cern.ch}

\author{Thomas Konstandin$^{(1)}$ and Stephan J. Huber$^{(2)}$ }

\email[]{Stephan.Huber@cern.ch}
\email[]{Konstand@kth.se}

\affiliation{$(1)\,$Department of Physics, Royal Institute of Technology (KTH),
        AlbaNova University Center,
        Roslagstullsbacken 11, 106 91 Stockholm, Sweden}

\affiliation{$(2)\,$ Theory Division, CERN,
        CH-1211, Geneva 23, Switzerland}

\date{\today}

\begin{abstract}
We present an algorithm to analyze numerically the bounce solution
of first-order phase transitions. Our approach is well suited to
treat phase transitions with several fields.  The algorithm
consists of two parts. In the first part the bounce solution without
damping is determined, in which case energy is conserved. In the second
part the  continuation to the physically relevant case with
damping is performed. The presented approach is numerically stable and easily implemented.
\end{abstract}

%\pacs{ 
%98.80.Cq,  % Particle-theory and field-theory models of the early Universe
%           % (including cosmic pancakes, cosmic strings, chaotic
%           % phenomena,
%           % inflationary universe, etc.)
%11.30.Er,  % Charge conjugation, parity, time reversal,
%           % and other discrete symmetries
%11.30.Fs   % Global symmetries (e.g., baryon number, lepton number)
%}

\maketitle

%
%%%%%%%%%%%%%%%%%%%%%%%%%%%%%%%%%%%%%%%%%%%%%%%%%%%%%%%%%%%%%%%%%%%%%%%%%%%%%%%
%  MAIN TEXT
%%%%%%%%%%%%%%%%%%%%%%%%%%%%%%%%%%%%%%%%%%%%%%%%%%%%%%%%%%%%%%%%%%%%%%%%%%%%%%%
%

\section{Introduction}

The problem of calculating properties of a first-order phase
transition from a false vacuum to an energetically lower vacuum
appears in different contexts of physics, as for example condensed
matter physics~\cite{Langer:1967ax,Langer:1969bc}, particle
physics~\cite{cole1,cole2}, and
cosmology~\cite{Affleck:1980ac,Linde:1977mm,Linde:1981zj}. In a
cosmological context, one of the interesting features of a first-order
phase transition is that it can proceed by bubble nucleation. The
subsequently expanding bubbles can be utilized as a source that drives
the hot plasma in the early Universe out of equilibrium. This is in
turn one of the Sakharov conditions, the requirements for a viable
baryogenesis mechanism.  Employing the electroweak phase transition,
the corresponding baryogenesis mechanism was suggested in
Ref.~\cite{Kuzmin:1985mm} and named electroweak baryogenesis.  Even
though this scenario of the generation of the baryon asymmetry of the
Universe (BAU) is excluded in the Standard Model (SM) because of a
lack of CP violation and the cross-over nature of the phase
transition, this possibility is still open in the Minimal
Supersymmetric Standard Model (MSSM)~\cite{Carena:2002ss,
Carena:2000id, Konstandin:2005cd}. An essential input parameter of the
determination of the produced BAU is the wall shape of the expanding
bubbles. While in the SM the phase transition is described by a single
{\it vev} of the Higgs field $v$, there are already two dynamical
fields during the phase transition in the MSSM, namely two {\it vevs}, 
$v_1$ and $v_2$, or equivalently, $v=\sqrt{v_1^2 + v_2^2}$ and
$\tan(\beta) = v_2/v_1$.  Notice that the initial value of $\beta$
cannot be inferred without studying the dynamics of the phase
transition, since $v$ vanishes in the symmetric phase.  This is
especially important, because some CP violating sources are
proportional to the change of $\beta$ during the phase 
transition.

 Examining extensions of the MSSM, one has eventually to deal with more
than two scalar fields. This is the case {\it e.g.} in the MSSM
extended by an additional singlet~\cite{Davies:1996qn,
Bastero-Gil:2000bw,Menon:2004wv}, which provides the interesting
feature of transitional CP violation~\cite{Huber:2000mg}. Further
examples are provided by models that equip this singlet with a
$U(1)^\prime$ gauge symmetry~\cite{Kang:2004pp} or by the
SM with two Higgs doublets~\cite{Cline:1995dg,Cline:1996mg}.

The theoretical basis for the determination of the tunneling rates in
particle physics and the explicit solution for one scalar field was
given in~\cite{cole1,cole2}.  However to calculate the properties in
the case of several scalar fields is analytically not accessible, and
even numerically a non-trivial task. We present a numerical solution to
this problem.

\section{The Problem}

\begin{figure}[h]
\begin{center}
\epsfig{file=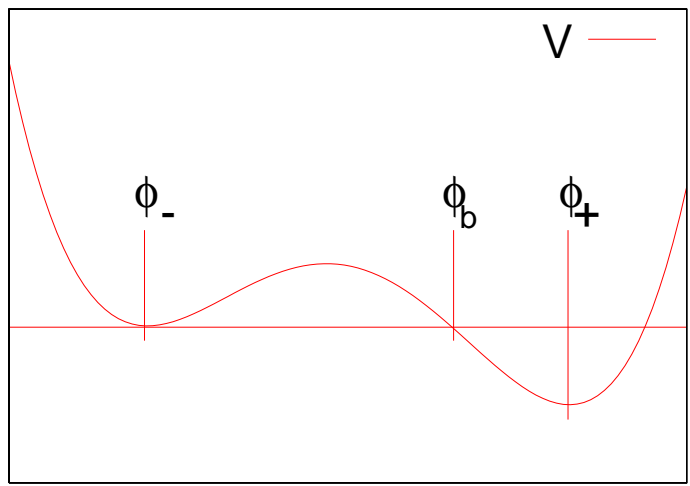, width=4.0 in}
\end{center}
\vskip -0.2in
\caption[Fig:fig_eta]{%
\label{fig_pot}
\small
The potential for a theory with a false vacuum.
}
\end{figure}

Before the numerical treatment of the multi field case is discussed,
we will briefly review the results of~\cite{cole1,cole2} and clarify
our notation. Given the Lagrangian
\be
{\cal L} = \frac12 \partial_\mu \phi \partial^\mu \phi - V(\phi), \label{lagrange}
\ee
the transition probability per unit time and unit volume from a false vacuum to 
an energetically lower vacuum
is, in the semi-classical approximation, given by 
\be
\Gamma = \frac{S[\bar \phi]^2}{4 \pi^2 \hbar^2} \, \exp({-S[\bar \phi]/\hbar})
\left| \frac{\det^\prime[-\partial^2 + V^{''}(\bar \phi)]}{\det[-\partial^2 + V^{''}(0)]} 
\right|^{-\frac12} \times (1 + O(\hbar)). \label{gamma_V}
\ee
The primed determinant indicates that the zero-modes, corresponding to 
translation invariance, have been omitted, leading to the factor $S[\bar \phi]^2$
in the formula.
The function $\bar \phi$ is the classical Euclidean solution of the equations of motion
derived from the Lagrangian in \eq~(\ref{lagrange}) with {\it inverted} potential
and the boundary conditions
\be
\lim_{\tau \to \infty} \bar \phi(\tau, \vec x) =\phi_-, \label{bounce1}
\ee
where $\phi_-$ denotes the false vacuum and 
\be
\partial_\tau \bar\phi(0, \vec x) =0.\label{bounce2}
\ee
In addition the solution should be non-trivial in the sense that
$\bar\phi(0, \vec x) = \phi_b$ is close to the new vacuum. The
corresponding solution is usually called bounce configuration and 
a typical potential is illustrated in Fig.~\ref{fig_pot}.

The present work aims at determining numerically the bounce
configuration $\bar\phi$ in theories with several scalar fields.  Thus,
the parameters that are inferred about the phase transition rely on the
semi-classical expansion in Eq.~(\ref{gamma_V}), which can break down,
{\it e.g.} for very weak first-order phase
transitions~\cite{Strumia:1998qq}.

In addition, the potential is treated as static and it is assumed
that the phase transition occurs at  some fixed
temperature. Issues concerning the timing of the phase transition can
be decided on by comparing the nucleation rate with the corresponding
 cooling time scales, such as the expansion rate of the
Universe. This includes, for instance, the questions to know if the vacuum resides
in a metastable state until today or if the phase transition is
performed before the Universe cools down to a temperature where only a
second-order phase transition is possible~\cite{and_hall}.

A crucial question in this
context, relevant to electroweak baryogenesis, is to see if the phase
transition is strong enough to avoid the washout of the generated
baryon asymmetry~\cite{Shaposhnikov:1986jp}. Fortunately, it turns
out that the  perturbative effective potential even slightly underestimates the
strength of the phase transition in the MSSM~\cite{Laine:1998vn, Laine:1998qk}. 

Finally, for certain physical problems the knowledge of
the prefactor in the decay rate of Eq.~(\ref{gamma_V}) might be necessary,
which we do not attempt to achieve. For a recent development and a
review on the topic, see Refs.~\cite{Dunne:2005rt, Baacke:2003uw}. Since
all these considerations are model-dependent, we will focus in the
following on the numerical determination of the bounce solution.

The solution for one scalar field in the thin-wall regime can be
analytically determined by using the saddle-point method~\cite{cole1,
cole2}.   Using the
fact that there exists a solution that is invariant under Euclidean
four-dimensional rotations, and hence that only depends on the combination
\be
\rho = \sqrt{\tau^2 + x^2},
\ee
the corresponding action and equation of motion (EoM) read
\be
S[\phi(\rho)] = 2\pi^2 \int_0^\infty d\rho \,\rho^3 \left[ 
\frac12 \left(\frac{d\phi}{d\rho} \right)^2 + V(\phi) \right] 
\ee
and
\be
\frac{d^2\phi}{d\rho^2} + \frac3\rho \frac{d\phi}{d\rho} = 
\frac{d}{d\phi}V(\phi). \label{EoM_old}
\ee 
Numerically, in the case with one field, this solution can be easily
found by the ``overshooting--/undershooting'' method. Varying the initial
point $\phi_b$ and testing if the configuration after integration of
the EoM over-- or undershoots the position $\phi_-$, the bounce
point $\phi_b$ and the corresponding configuration can be
determined. Clearly this method cannot be generalized to the
multi field case, since starting at $\phi_+$ there is no viable
strategy to find the bounce point $\phi_b$.

For tunneling in a thermal system, the $O(4)$ symmetry is replaced by
an $O(3)$ symmetry and the required solution is constant in 
time~\cite{Affleck:1980ac,Linde:1977mm,Linde:1981zj}.
The corresponding action and EoM read
\be
S[\phi(\rho)] = 4\pi \int_0^\infty  d\rho \, \rho^2 \left[ 
\frac12 \left(\frac{d\phi}{d\rho} \right)^2 + V(\phi,T) \right] 
\ee
and
\be
\frac{d^2\phi}{d\rho^2} + \frac2\rho \frac{d\phi}{d\rho} = 
\frac{d}{d\phi}V(\phi, T), \label{EoM_old2}
\ee 
and the ``overshooting--/undershooting'' method can be utilized in a similar fashion.
Generalizing these two cases and allowing for $n$ scalar fields,
the corresponding action yields
\be
\tilde S_\alpha[\phi(\rho)] = \int_0^\infty d\rho \, \rho^{(\alpha-1)} \left[ \frac12 
\left(\frac{d\phi}{d\rho} \right)^2 + V(\phi) \right], \label{action}
\ee
where $\alpha=3$ ($\alpha=4$) in the temperature (vacuum) case, $\phi$
denotes a vector with $n$ entries, and a constant
prefactor of the action is neglected.  The $n$ equations of motion read
\be
\frac{d^2\phi}{d\rho^2} + \frac{(\alpha - 1)}{\rho} \frac{d\phi}{d \rho} 
= \nabla V(\phi)
\label{EoM_alpha}
\ee 
and can be interpreted as a classical particle moving in the
$n$-dimensional inverted potential with a time-dependent damping term.
 In the thin-wall regime, where the maxima are almost degenerate,
$V(\phi_+) \approx V(\phi_-)$, one can determine the action analogously to
that in Ref.~\cite{Linde:1977mm},  as
\be
\tilde S_\alpha[\bar\rho] =-\frac{ \bar\rho^\alpha \, \delta V }{\alpha} + 
\bar\rho^{(\alpha-1)} \tilde S_1,	
\ee
where $\delta V$ denotes the potential difference between the two
maxima and $\bar \rho$ the radius of the bubble. Minimization of the
action leads to the radius of the bubble
\bea
\bar\rho = (\alpha-1) \, \frac{\tilde S_1}{\delta V}, \label{r_0_eq}
\eea
and yields the following estimate of the action in the thin-wall regime
\be
\tilde S_\alpha[\phi(\rho)] = 
\frac{(\alpha-1)^{(\alpha-1)} \, {\tilde S_1^\alpha}} { \alpha \, \delta V^{\alpha-1}}.
\label{action_est}
\ee

The problem is now to determine the bounce solution apart from the
thin-wall regime. Even with only two fields, it is a challenging task
to find numerically the position $\phi_b$ where the fields bounce.
One possibility is, starting at $\phi_-$, to ``scan'' into different
directions. For two fields this seems to be feasible, but it clearly fails
for more than two fields.

Alternatively, one could try to obtain the bounce solution by
minimizing the action. However, it was already pointed out by Coleman
that the bounce is not a minimum of the action but a saddle
point~\cite{cole3}.  In the following paragraphs we will briefly
review former techniques to solve this problem, before we discuss our
approach in Sections~\ref{our_app} to~\ref{trans}.

The first improvement was made in Ref.~\cite{claudson}, where 
a tunneling amplitude in supergravity is discussed. The
authors identified the direction of the negative eigenvalue of the
action to be the scaling mode of the field $\phi$. Accordingly, they
minimized the action in several iterations and maximized it with
respect to scaling, simultaneously. However, to determine the
directions orthogonal to scaling is numerically a difficult task. The
authors reported errors of $3$--$5\%$ in the solution, which is due to the
fact that the system converges to a point where scaling and
minimizing balance, and not to the real bounce solution.

This weakness was criticized in Ref.~\cite{kusenko}, where another paradigm was followed, 
the so-called improved-action approach. The direction of the negative eigenvalue 
was lifted in the vicinity of the bounce by using a generalized 
version of the viral theorem. To implement this procedure successfully, one needs 
a configuration that is already close to the bounce. 
%In addition, the improved action is 
%not differentiable and one relies on random deviation algorithms
%to minimizing the action
%and cannot use more sophisticated
%algorithms as e.g. Newton's  method that show better convergence properties.

Further approaches are based on minimizing an action that consists of
the squared equations of motion~\cite{seco, john}. This method also
depends on a good anticipation of the bounce. In addition, it was
reported in Ref.~\cite{john} that in certain cases the implementation is
numerically not stable in the sense that convergence to the bounce is
not guaranteed and an oscillatory behavior and
unphysical solutions can evolve.

The last two methods are not well suited to problems with more than
two fields, since a good anticipation of the bounce is in this case
usually not possible.

Another method based on performing a certain order of gradient descent
and ascent steps was applied in Ref.~\cite{Cline:1999wi}. In this
algorithm, the ascent steps ensure that the configuration approaches a
local extremum and not just a local minimum.  Additionally, a second
algorithm based on a combination of the shooting and the
gradient-descent method was presented in this work. This second
algorithm is slightly more technically involved than our proposal,
since it uses spline interpolation.  These two algorithms have a
similar range of validity as our approach, but they are slower,
so it is more numerically expensive to achieve high accuracy.

To find the bounce solution, we split the
problem into two parts.  First, we determine the solution without
damping ($\alpha=1$). Subsequently, we perform a smooth transition to
the damped case of interest ($\alpha=3$ or $\alpha=4$).
Neglecting the ``damping'' term is essential in the first part,
since the corresponding approach is based on the assumption that 
energy is conserved and the bounce point $\phi_b$
is on the submanifold with potential equal to $V(\phi_-)$ 
(as already indicated in Fig. \ref{fig_pot}).

\section{Improved Potential Approach \label{our_app}}

In this section we will present our approach to find the bounce solution
in the undamped case, $\alpha=1$.

First, notice that, even in the undamped case, the initial problem remains and 
the bounce configuration is only a saddle point of the action.
 Intuitively, this can be understood as follows. 
The undamped case can be interpreted as a particle moving freely in a potential.
To minimize the action, the system tries to avoid valleys, and hence moves
fast in them. If
the system has a long time to evolve, the global minimum of the
action is given by a field configuration that shoots up to the maximum
of the inverted potential $\phi_+$ and stays there as long as
possible, only restricted by the boundary conditions. Thus, the bounce
solution cannot be a global minimum of the action.

Furthermore, close to the bounce solution 
that fulfills the boundary conditions: 
\be
\phi(0)=\phi_b, \quad \phi(\infty)=\phi_-, 
\ee
the action can be reduced by increasing $\partial_\rho \phi(0)$ in the direction
of maximal slope (towards $\phi_+$).  Hence the bounce configuration
constitutes a saddle point of the action, even though the ``damping''
term is neglected. Notice that in the case of degenerate minima the
negative eigenvalue corresponding to the instability of the saddle
point~\cite{claudson} becomes zero and reflects the invariance under
translation in the coordinate $\rho$.  The solution of the degenerate
case without damping can be easily obtained by minimization of the
action, and the subtlety of instabilities appears only if the minima
are non-degenerate.

 Our starting point for the determination of the bounce solution is 
the configuration restricted by the following boundary conditions at the 
maxima of the inverted potential:
\be
\phi(0)=\phi_+, \quad \phi(T)=\phi_-, \label{bc_1}
\ee
where $T$ is a time scale that is sent to $\infty$ iteratively.
Given that $\phi_+$ is the global maximum of the inverted potential, 
there exists a solution with these boundary conditions that indeed is 
a minimum of the action in \eq~(\ref{action}) and can
therefore be found by simply discretizing the field configuration and 
minimizing the action using the Newton method.

In the case of non-degenerate minima the resulting configuration is
not the bounce. Our goal is to modify the potential in such a way that
the minimization of the standard action with modified potential leads
first to a good prediction of $\phi_b$ and finally to the bounce
configuration.  With respect to the original problem, only
changes in the potential are performed, and an interpretation in terms of a
classically moving particle is viable throughout the present discussion.

In a first step the potential is changed by performing the following 
transformation:
\be
U_\epsilon(\phi) = \frac{V(\phi)-V(\phi_-)}{2} 
	+ \sqrt{\frac{(V(\phi)-V(\phi_-))^2}{4} + \epsilon^2}.
\ee
For finite $\epsilon$, this modifies the potential in a differentiable
way and in the limit $\epsilon \to 0$ transforms the potential in such
a way that it is unchanged (besides a constant) for {$V\geqslant V(\phi_-)$}
and develops a constant plateau in the region with $V \leqslant
V(\phi_-)$.

In the simultaneous limits
\be
\epsilon \to 0, \quad T \to \infty, \label{limit}
\ee
the minimum of the modified action will determine the bounce
solution. Here, the uniqueness of the minimum of the action is
assumed, even though in some cases this is not
valid~\cite{Kusenko:1995bw}.  To show that this minimum determines the
bounce configuration, the configuration that corresponds to the
minimum of the modified action is split into two parts. The first
part consists of the path in the plateau, while the second part is
outside the plateau, and both are connected by a point
$\phi_e=\phi(t_e)$
 that is defined by $V(\phi_e)=V(\phi_-)$.
The path in the plateau does not contribute to the
action in the limit $t_e \to \infty$, $\epsilon
\to 0$, and hence does not restrict the point $\phi_e$ or the path
outside.  In the limit $\epsilon \to 0$, the path outside the plateau
fulfills the Euler--Lagrange equations in the original potential
$V(\phi)$.  Using the fact that, in this limit, $U_\epsilon(\phi_-)$ and
$U_\epsilon(\phi_e)$ both approach the global maximum
$U_\epsilon(\phi_+)$ of the inverted potential, leads to the fact that
\be
\partial_\rho \phi(t_e) \to 0, \quad \partial_\rho \phi(T) \to 0 
\ee
in the limit $(T-t_e) \to \infty$. Hence, 
the path outside the plateau defines a bounce solution in this limit
according to Eqs.~(\ref{bounce1}) and (\ref{bounce2}). 

For finite $\epsilon$ and $T$, the field configuration will move very
slowly in the plateau from $\phi_+$ to $\phi_e$ (a point at a distance
$O(\epsilon, 1/T)$ to $\phi_b$) and then move from $\phi_e$ to
$\phi_-$, much as the bounce solution. This way, a fairly good
estimate of the bounce configuration and of $\phi_b$ is obtained even
for finite $\epsilon$ and $T$.  On the other hand, this configuration
still misses one typical feature of the bounce, namely the exponential
behavior of $\bar\phi$ close to $\phi_-$. Of course, this feature
appears in the limit of Eq.~(\ref{limit}), but the numerical stability
and performance of the algorithm can be improved by an additional
change of the potential. The reason for the absence of the exponential
behavior is that, since the global minimum of the potential is still
close to $\phi_+$, the configuration spends most of the time in the
plateau. To avoid this effect, another term can be added to the
potential, of the form
\bea
\tilde U_\epsilon(\phi) &=& U_\epsilon(\phi) + \Delta U_\epsilon(\phi) \\
\Delta U_\epsilon(\phi) &=&
- 2 \epsilon \frac{|\phi-\phi_-|^3}{|\phi_+-\phi_-|^3} 
+ 3 \epsilon \frac{|\phi-\phi_-|^2}{|\phi_+-\phi_-|^2}. \label{lift}
\eea
This additional contribution is designed to lift the potential so that 
\be
0 \lesssim \tilde U_\epsilon(\phi_+) - \tilde U_\epsilon(\phi_-) \lesssim O(\epsilon)
\ee
and the corresponding solution shows the exponential behavior close to
$\phi_-$. This lifting is in principle not necessary, but it improves
the performance of the algorithm drastically.  Initially, a reasonable
choice for $\epsilon$ is
\be
\epsilon_{\textrm{\small init}} = V(\phi_+) - V(\phi_-),
\ee
 while for $T$ a time that approximately corresponds to 20 times the
thickness of the wall is
appropriate. The wall thickness can be anticipated by determining the
local minimum of the inverted potential and $|\phi_- - \phi_+|$
\be
T_{\textrm{\small init}} = 20 \times \frac{|\phi_- - \phi_+|}{\sqrt{8|V_\textrm{\small max} - V_\textrm{\small min}|} }.  
\ee

By the procedure described above one obtains a good estimate $\phi_e$
of the bounce point $\phi_b$ and a configuration close to the desired
bounce configuration. The convergence can be further improved by
truncating the part of the configuration in the plateau.
The boundary conditions are accordingly changed to
\be
\phi(0)=\phi_e, \quad \phi(T)=\phi_- \label{bc_2},
\ee
where $\phi_e$ initially denotes the position that corresponds to the
point with $V(\phi_e)=V(\phi_-)$ on the field configuration that was
obtained by taking the previously described steps.  In the next
minimization procedure, $\phi_e$ is allowed to move freely on the
submanifold with $V(\phi_e)=V(\phi_-)$ so that $\phi_e \to \phi_b$
is ensured.  While minimizing the action with the potential $\tilde
U_\epsilon(\phi)$, $\epsilon$ can be iteratively sent to zero until
the numerical uncertainties inhibit further improvement of the bounce
configuration.

To truncate the configuration in the plateau is helpful for two
reasons. First, the energy is conserved and is exactly $V(\phi_-)$ for
the bounce solution.  The motion in the plateau from $\phi_e$ to
$\phi_+$ contributes at least $(\phi_e-\phi_+)^2/2 T^2$ to the kinetic
energy, which requires very large $T$ for accurate results. This would
lead to large lattices, provided the physical lattice spacing is kept
constant. This effect is avoided by starting at $\phi_e$. Secondly,
since $\tilde U_\epsilon(\phi)$ is very flat in the plateau for
$\epsilon \to 0$, the convergence of the Newton method is very poor in
this region. Starting at $\phi_e$, the field never enters the plateau.

Even though the changes outside the plateau are only marginal for
$\epsilon \to 0$, it is clear, from classical reasoning, that this
procedure stabilizes the unstable mode of the original action.

\section{Numerical Results of the Undamped Case}

As a first example, the degenerate case is demonstrated. Our approach is
applied to the following potential with one scalar field and
degenerate minima
\begin{figure}[t]
\begin{center}
\epsfig{file=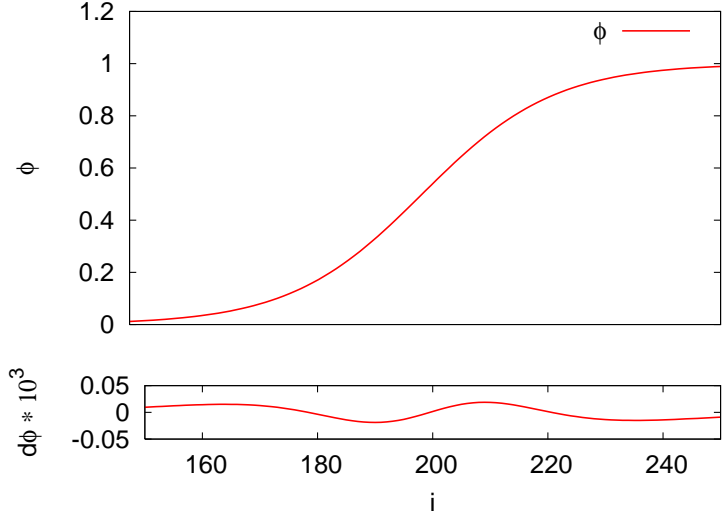, width=4.0 in}
\end{center}
\vskip -0.2in
\caption[Fig:fig_eta]{%
\label{fig_degen_phi}
\small
 The bounce configuration in the one-dimensional degenerate case.
The deviations from the exact analytic result are $O(10^{-4})$.
}
\end{figure}
\be
V(\phi) = 2 (\phi - 1)^2 \phi^2.
\ee
The resulting configuration and the deviation from the exact analytic result are plotted in
Fig.~\ref{fig_degen_phi}.
\begin{figure}[t]
\begin{center}
\epsfig{file=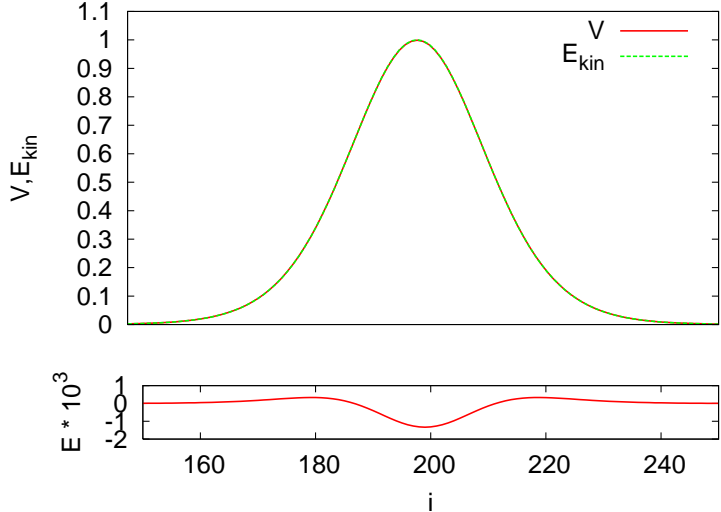, width=4.0 in}
\end{center}
\vskip -0.2in
\caption[Fig:fig_eta]{%
\label{fig_degen_ene}
\small
 Potential $V$ and kinetic energy $E_\textrm{kin}$ of the
 one-dimensional configuration. The total energy
 $E=E_{\textrm{kin}}-V$ is conserved up to $0.1\%$.  }
\end{figure}
The deviation from the exact solution is $O(10^{-4})$ and
mostly due to the finite lattice spacing. Another test is provided by
the kinetic energy that should vanish in the limits $\rho \to 0$
and $\rho \to T$.  This was not enforced by the boundary conditions
since the start and end-point of the configuration are used as boundary
conditions of the discretized action, as given in Eq.~(\ref{bc_1}) or
Eq.~(\ref{bc_2}). In addition one can analyze the conservation of
energy.  These tests are displayed in Fig. \ref{fig_degen_ene}.

\begin{figure}[t]
\begin{center}
\epsfig{file=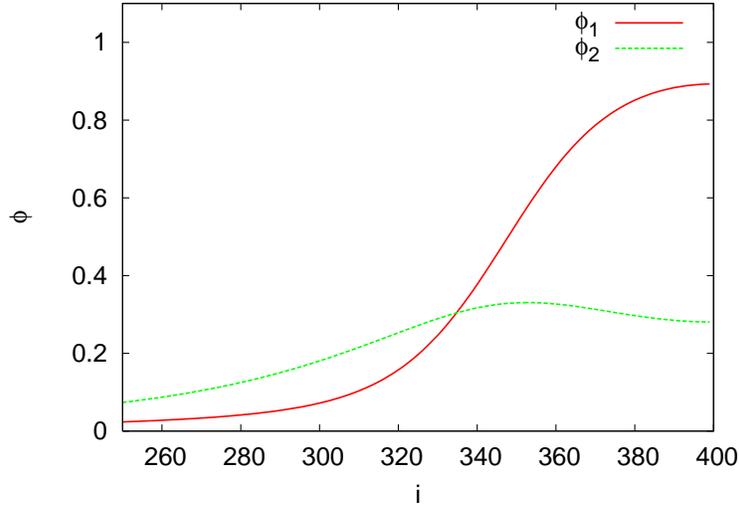, width=4.0 in}
\end{center}
\vskip -0.2in
\caption[Fig:fig_eta]{%
\label{fig_ndegen_phi}
\small
 The bounce configuration in the two-dimensional non-degenerate case.
The exponentially decreasing tail $[0,250]$ has been cut off.
}
\end{figure}

\begin{figure}[t]
\begin{center}
\epsfig{file=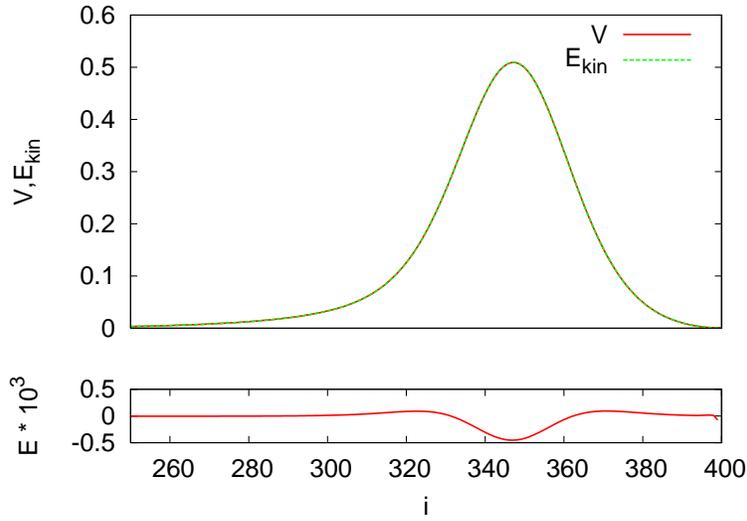, width=4.0 in}
\end{center}
\vskip -0.2in
\caption[Fig:fig_eta]{%
\label{fig_ndegen_ene}
\small
 Potential $V$ and kinetic energy $E_\textrm{kin}$ of the
 two-dimensional configuration. The total energy $E=E_\textrm{kin} -
 V$ is conserved up to $0.1\%$. The exponentially decreasing tail
 $[0,250]$ has again been cut off.  }
\end{figure}

In the second example, a non-degenerate potential for the multi scalar
case is analyzed, namely
\be
V(\phi) = 16 (\phi_1 - 1)^2 \phi_1^2 + 2\,\phi_2^2 + c_1 \, \phi_1 +
c_2 \, \phi_2 \phi_1 (\phi_1-1).
\ee
The third term dissolves the degeneracy of the minima and is chosen
to be rather small ($c_1=-0.1$). The last term gives a non-trivial
dynamics in the $\phi_2$ direction ($c_2=8.0$).  The numerical results
are depicted in Figs.~\ref{fig_ndegen_phi} and~\ref{fig_ndegen_ene}.
The results show the expected exponential behavior close to $\phi_-$
and a parabolic shape near $\phi_b$. The energy is again conserved up to 
the per mille level. Especially the parabolic shape 
with $\partial_\rho \phi = 0$ at the bounce point $\phi_b$ 
in Fig. \ref{fig_ndegen_phi} is a clear indication
of the high precision of our numerical approach, since because of the lift 
of the potential in \eq~(\ref{lift}), the kinetic energy is there $O(\epsilon)$. 

Note that the shape of the field $\phi_2$ is not of the usual
{\it hyperbolic tangent} form, and not even monotonically increasing; hence
one cannot obtain a good estimate of the bounce configuration by
minimizing the action with respect to a specific {\it ansatz}, as was
done in Ref.~\cite{john}.

\section{ Continuation to the Damped Case\label{trans}}

To perform the  continuation from the undamped case ($\alpha=1$) to the
damped case, a numerical method to solve the discretized version of
Eq.~(\ref{EoM_alpha}) with the boundary conditions as given in
Eqs.~(\ref{bounce1}) and (\ref{bounce2}) is needed.  In the present
work the following discretized form of the EoM has been 
used: 
\be
\frac{\phi_{i+1} - 2 \phi_i + \phi_{i-1}}{ d\rho^2}
+ \frac{\alpha-1}{(i + \Delta i) d\rho } \frac{\phi_{i+1} - \phi_{i-1} }{ 2\, d\rho }
= \nabla V(\phi_i), \quad i \in [1,n-1], \label{disc_eom}
\ee
with the boundary conditions
\be
\phi_0=\phi_1, \quad \phi_n=\phi_-,
\ee
where $\phi_i$ constitutes the one-dimensional lattice $(i \in
[0,n])$. 
This discretization has the virtue to be derived from the energy functional 
\be
H = \sum_i i^2 V(\phi_i) + \sum_i (i^2+i) \frac{(\phi_{i+1}-\phi_i)^2}{2 d\rho^2}
\ee
and reproduces the continuum energy with an error of $O(d\rho^2)$.

The offset parameter $\Delta i$ has been introduced to
improve the performance of the algorithm, since an initial value of
$\Delta i=0$ can easily lead to a pathological behavior close to
$i=0$. Finally, this parameter has to be sent to zero. 

One way to obtain a solution of this system, is to use the method of
squared equations of motion, as was done in Refs.~\cite{seco,
john}. However, since the algorithm in the last sections provides
fairly good estimates $\tilde\phi_i$ of the solution, we pursue
another possibility. Using the Taylor expansion of the gradient field
\be
\nabla V(\phi_i) \approx 	\nabla V(\tilde\phi_i)
 + (\phi_i - \tilde\phi_i) \cdot \nabla (\nabla V(\tilde\phi_i)),
\ee
we can linearize the system of equations and iteratively solve the
equations by inversion of the linear system.  Since the matrix
corresponding to this system of equations is band diagonal, its
inversion is easily performed. The original system of equations in
Eq.~(\ref{disc_eom}) is then solved iteratively by using the
linearized system.  Notice that the original potential is used and not
the modified potential that has been utilized in the minimization of
the action in the last section.

This method has been applied to the non-degenerate example of the last
section and the results are briefly displayed in the
following. Starting point of the  continuation are the damping parameter
$\alpha=1$, an offset $\Delta i =300$, a time scale $dt=0.0158$, and a
lattice of size $n=400$. As initial configuration, the one
generated by the improved-potential method is used, as discussed in the
last section.

\begin{figure}[t]
\begin{center}
\epsfig{file=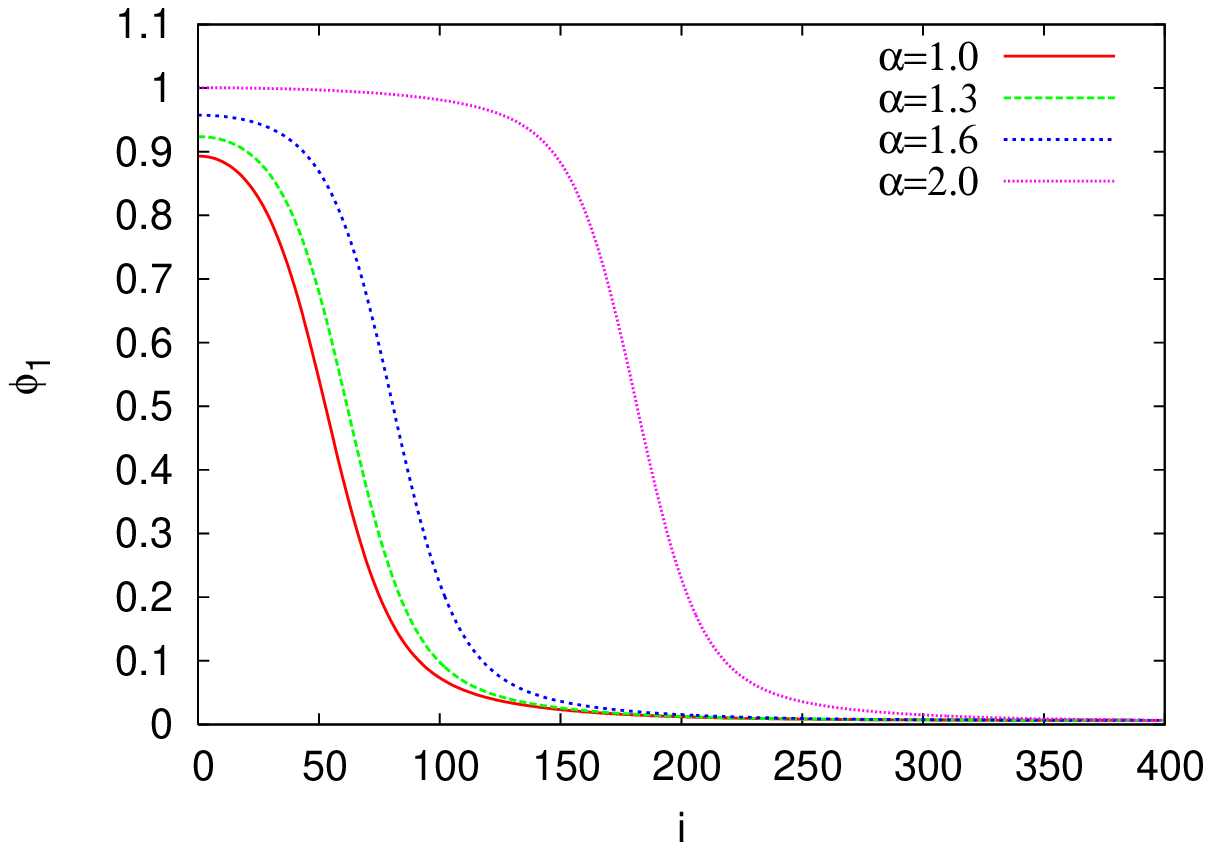, width=3.0 in}
\epsfig{file=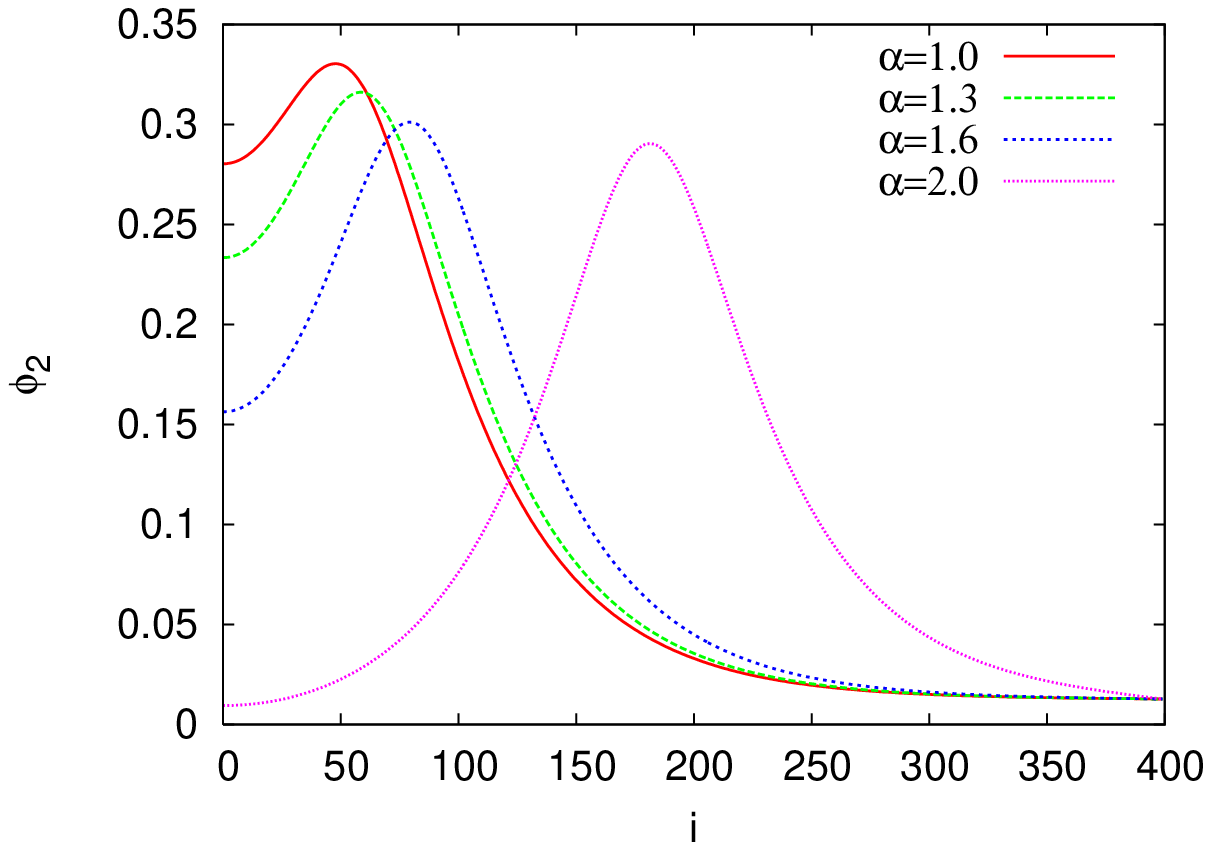, width=3.0 in}
\end{center}
\vskip -0.2in
\caption[Fig:fig_eta]{%
\label{fig_transition}
\small
 The fields $\phi_1$ and $\phi_2$ during the  continuation for several values of
$\alpha$ and $\Delta i=300$.  }
\end{figure}

\begin{figure}[t]
\begin{center}
\epsfig{file=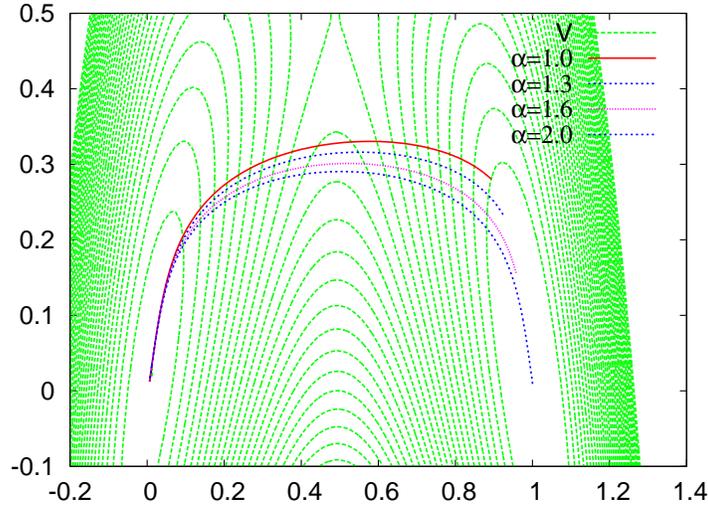, width=4.0 in}
\end{center}
\vskip -0.2in
\caption[Fig:fig_eta]{%
\label{fig_contour}
\small
Contour plot of the potential and several trajectories.
}
\end{figure}

The parameter $\alpha$ is increased by steps of size $\delta \alpha= 5
\times 10^{-3}$ and the linear system in
Eq.~(\ref{disc_eom}) is solved iteratively for each parameter set and
some obtained configurations are plotted in
Fig.~\ref{fig_transition}. The contour of the potential and several
trajectories are depicted in Fig.~\ref{fig_contour}.

As soon as the size of the bubble
\be
\bar\rho = (\alpha-1) \, \frac{\tilde S_1}{\delta V} - \Delta i \, dt, \label{r_0_eq_num	}
\ee
is larger than the wall thickness, an increase in $\alpha$ or a
decrease in $\Delta i$ results almost only in a translation of the
configuration. If this translation is anticipated, a faster  continuation
in the parameters $\alpha$ and $\Delta i$ becomes possible.

Since solving the linearized system of equations is not very time
consuming, the accuracy of the result is in principle only limited by
the finite size of the lattice and the accuracy of the numerical
representation. This can be demonstrated by increasing the number of
lattice sites and the results are summarized in
Table~\ref{tab_results}.  The relative error in the action $S_3$
behaves as expected as $O(\delta\rho^2)$ and reaches approximately $7
\times 10^{-6}$ for the largest lattice that was used. Errors of
$O(\delta\rho^4)$ can be obtained by extrapolation of the action to
vanishing $\delta\rho$.  In comparison, using the method of squared
EoM or one of the algorithms of Ref.~\cite{Cline:1999wi} is much more time
consuming and leads to inferior accuracy.
\begin{table}
\begin{center}
\begin{tabular}{|c|c|c|}
\hline 
lattice sites & $S_3$ & $\Delta S_3$ \\
\hline 
100 &  725.819 &   \\
\hline 
200 & 726.957 & 1.137 \\
\hline 
400 & 727.239 & 0.282 \\
\hline 
800 & 727.309 & 0.070 \\
\hline 
1600 & 727.326 & 0.017 \\
\hline 
\end{tabular}
\end{center}
\caption{
\label{tab_results}
The action $S_3$ for different numbers of lattice sites. The change in the action
$\Delta S_3$ behaves as $O(\delta \rho^2)$. The error of the largest lattice can 
be estimated to be $5 \times 10^{-3}$ what corresponds to a relative error of 
$7 \times 10^{-6}$.
}
\end{table}

\section{Conclusion\label{concl}} 

We presented a numerically stable and easily implemented algorithm to
determine properties of phase transitions with several scalar fields. 
We demonstrated our approach in the simplest
one-dimensional and in a non-trivial two-dimensional example. The
approach we suggest converges with high precision to the bounce
configuration.  During the minimization neither oscillatory nor
unphysical solutions have been encountered, in contrast to the approach
followed in Ref.~\cite{john}.

Our approach was based on a procedure consisting of two
steps. First, we determined the bounce solution in the undamped
case, using a novel algorithm. Secondly, a smooth  continuation was
performed to the physical case, turning on the damping term in the
EoM.  Note that the sole purpose of the undamped
solution is to serve as an initial configuration for the  continuation.
In contrast to the configuration without damping, the damped
configuration usually shares more characteristics with the typical
solution in the degenerate case, namely a bubble size much larger than
the bubble wall thickness and an exponential behavior close to both
maxima. Hence the action determined in the thin-wall regime as
given in Eq.~(\ref{action_est}) is typically more accurate than using the
configuration of the undamped case in the action.

The main advantage of our approach, besides the very high accuracy, is the
large range of applicability. The convergence of our algorithm does
not rely on a good anticipation of the bounce and is hence well suited
to treat multi dimensional phase transition problems.  In a
forthcoming publication the method we presented will be used to
analyze the electroweak phase transition in the
nMSSM~\cite{forthcomming}.

\section*{Acknowledgments}
         
We would like to thank M. Seco for useful discussions and
T. H{\"a}llgren for commenting on the manuscript. 
Additionally, we would like to thank G. Moore for 
discussing with us some details of the algorithms presented 
in Ref.~\cite{Cline:1999wi}.
This work was
supported by the Swedish Research Council (Vetenskapsr{\aa}det),
Contract No.~621-2001-1611.

\end{document}